\documentclass[%
 preprint,
superscriptaddress,
nofootinbib,
nobibnotes,
 amsmath,amssymb,
 aps,
prx,
longbibliography,
showkeys
]{revtex4-2}
\usepackage{graphicx}  
\usepackage{dcolumn}   
\usepackage{bm}        
\usepackage{amssymb}   
\usepackage{siunitx}
\usepackage[breaklinks=true,colorlinks,citecolor=blue,linkcolor=red,urlcolor=blue]{hyperref}
\usepackage{multirow}
\usepackage{xcolor}
\usepackage{amsmath}
\usepackage{hhline}
\usepackage{lipsum}

\usepackage{soul}
\usepackage{color} 

\usepackage{braket}
\usepackage{autobreak}

\usepackage{amsfonts}
\usepackage{amsmath}
\usepackage{slashed}
\usepackage{amsthm}
\usepackage{amssymb}
\usepackage[vietnamese, english]{babel}
\usepackage{color}
\definecolor{mycolor2}{rgb}{0.0, 0.0, 1.0}
 
\begin{document}

\title{Tripartite Topological Equivalence of Feature, Entanglement, and Wilson Loop Spectrum from Nested Feature Spectrum Topology}

\author{Yi-Chun Hung$\,^{\hyperlink{equal}{*}}$}
\affiliation{ Institute of Physics,\;Academia Sinica,\;Taipei,\;115201,\;Taiwan}
\affiliation{ Department of Physics,\;Northeastern\;University,\;Boston,\;Massachusetts\;02115,\;USA}
\affiliation{ Quantum Materials and Sensing Institute, Northeastern University, Burlington, Massachusetts 01803, USA}

\author{T. Tzen Ong$\,^{\hyperlink{equal}{*},\hyperlink{email1}{\dagger}}$}
\affiliation{ Institute of Physics,\;Academia Sinica,\;Taipei,\;115201,\;Taiwan}

\author{Hsin Lin$\,^{\hyperlink{email2}{\ddagger}}$}
\affiliation{ Institute of Physics,\;Academia Sinica,\;Taipei,\;115201,\;Taiwan}


\begin{abstract}
Topological phases of matter are traditionally characterized through symmetry-based classifications. In cases of symmetry breaking, the projected spectrum — obtained by projecting the ground state onto the eigenstates of a pertinent quantum observable, such as spin or orbital angular momentum — provides a clear method for classifying topological phases. This approach underpins well-known frameworks such as spin-resolved topology and feature spectrum topology. Here we introduce nested feature spectrum topology, in which projection operators are applied recursively to subsectors of the feature spectrum, generating a hierarchy of feature spectra. We uncover a fundamental tripartite equivalence among the topology of feature, the entanglement, and the Wilson loop spectra in non-interacting fermionic systems. This equivalence reveals that the feature spectrum encodes the entanglement between sectors of the quantum observable, such as the spin-up and spin-down states in spin-resolved topology. We further prove that spectral flow in the entanglement spectrum and the Wilson loop winding in the feature spectrum are equivalent manifestations of the feature-energy complementarity: the appearance of gapless spectral flow in either energy or projected spectra on the boundary. This complementarity refines the conventional bulk-boundary correspondence by demonstrating that topological boundary modes may persist in the feature spectrum even when energy spectra are gapped. Our results provide a deeper understanding and solid foundation for the origin of band topology in the feature spectrum.
\end{abstract}

\renewcommand{\thefootnote}{\fnsymbol{footnote}}
\footnotetext[1]{\hypertarget{equal}{These authors contributed equally.}}
\footnotetext[2]{\hypertarget{email1}{Contact author: \href{mailto:tzen.ong@gmail.com}{tzen.ong@gmail.com}}}
\footnotetext[3]{\hypertarget{email2}{Contact author: \href{mailto:nilnish@gmail.com}{nilnish@gmail.com}}}

\maketitle
\section{Introduction}
Ground state topology is a cornerstone of modern condensed matter physics. Clear theoretical understanding and predictions for topological materials in non-interacting solids have driven numerous advances and applications in condensed matter physics and material science \cite{RevModPhys.88.021004, RevModPhys.82.3045, RevModPhys.83.1057}, culminating in the discovery of symmetry-protected topological (SPT) phases \cite{PhysRevB.78.195424, PhysRevB.76.045302, Po2017, Bradlyn2017-pj, PhysRevX.7.041069}. However, in experimental realities, perturbations and interactions inevitably break the protecting symmetries, technically invalidating the standard SPT characterization. This raises a fundamental question: does the topological character of the ground state vanish the moment symmetry is broken, or does it persist in a more subtle form?

To address this, recent research has developed a framework known as \textbf{Feature Spectrum Topology} \cite{FeatureSpectrumTopology}, which encompasses related concepts like spin-resolved topology \cite{PhysRevB.80.125327, Lin2024, PhysRevLett.108.196806}. In this context, the ground state Hilbert subspace is partitioned into distinct sectors by the feature spectrum, the spectrum of $P\hat{O}P$, and the ground state topology is characterized by the band topology of these sectors. Here, $P$ is the projection operator to the ground state and $\hat{O}$ is the operator that characterizes the system's ``feature" of interest. Besides spin, such a feature of interest can be applied to any translationally-invariant operator, including mirror symmetry \cite{FeatureSpectrumTopology} and orbital angular momentum \cite{PhysRevB.111.195102, PhysRevLett.126.056601}. The key discovery of this framework is \textbf{feature-energy complementarity}: even when symmetry-breaking perturbations open a gap in the boundary energy spectrum, the bulk-boundary correspondence persists as a gapless spectral flow within the boundary feature spectrum. This suggests that the bulk-boundary correspondence is more robust than previously thought, surviving in the projected Hilbert subspace of the sectors in the feature spectrum \cite{FeatureSpectrumTopology, PhysRevB.109.155143}.

The entanglement spectrum has widely served as a parallel alternative to conventional topological invariants for probing ground state topology, as it captures the entanglement between spatial partitions of the ground-state wave-function. In the presence of protecting symmetries, gapless spectral flow in the entanglement spectrum is viewed as a proxy for gapless boundary modes \cite{PhysRevB.101.115140, PhysRevB.83.245132, brzezinska_2018, PhysRevLett.113.106801, PhysRevLett.104.130502, PhysRevB.87.035119, Lin2024, PhysRevLett.107.036601}; thereby revealing topological order. Analogously to spin-resolved topology, the entanglement spectrum of individual spin sectors has been employed to characterize systems with broken symmetries \cite{doi:10.7566/JPSJ.85.043706, PhysRevB.96.165139, doi:10.7566/JPSJ.83.113705, doi:10.7566/JPSJ.84.043703}. However, the critical question of the bulk-boundary correspondence for such a spin-resolved entanglement spectrum and its corresponding physical nature, as well as its extension to more general features, remains unaddressed. 

\begin{figure}[h]
\centering
\includegraphics[width=\linewidth]{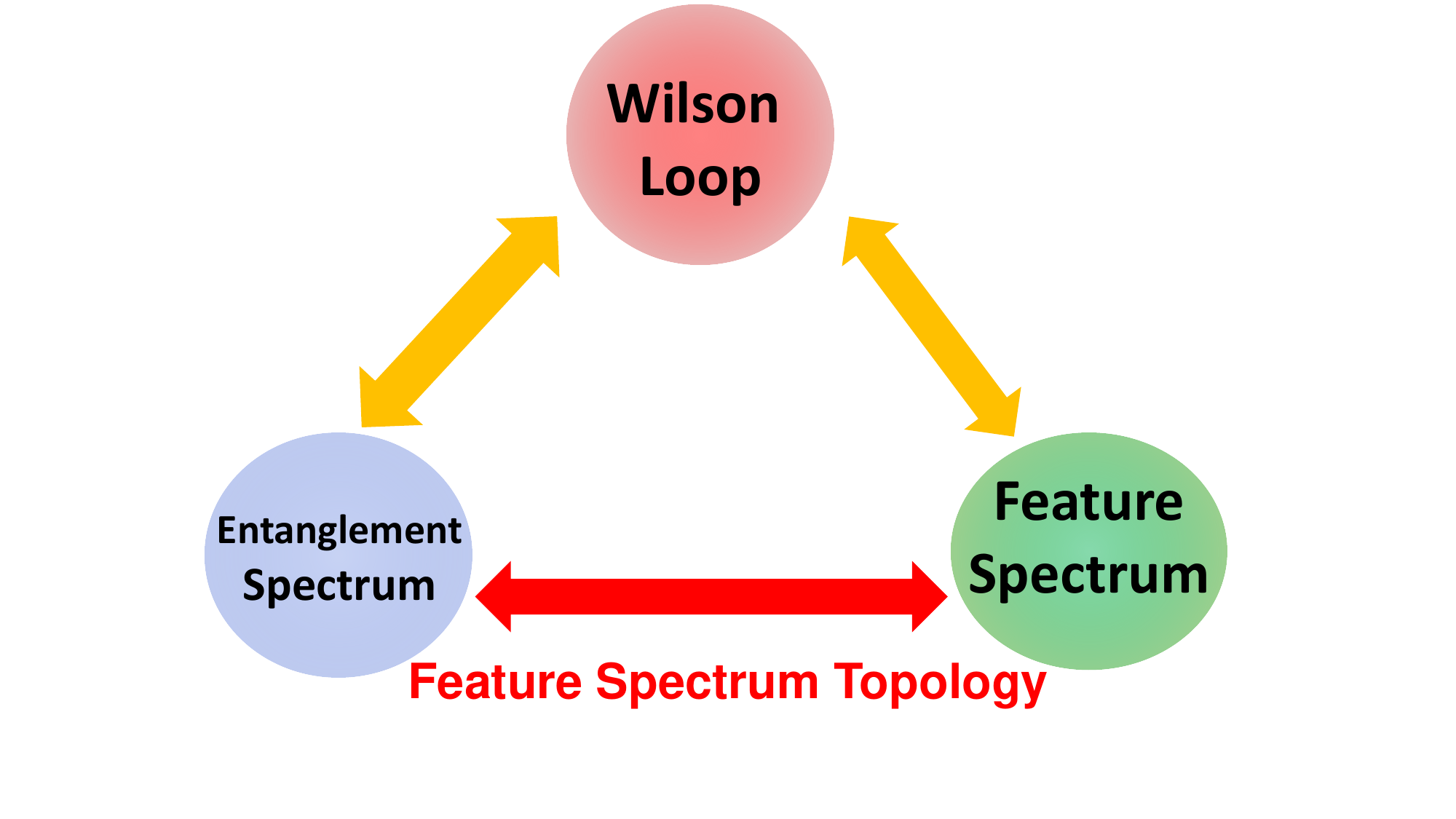}
\caption{The tripartite equivalence between the Wilson loop spectrum, spatially-resolved entanglement spectrum, and spatially-resolved feature spectrum has been extensively studied \cite{PhysRevLett.107.036601, J.Phys.A:Math.Gen.36.L205, PhysRevB.91.085119}. Previous research has elucidated the connections between the Wilson loop spectrum and both the spatially-resolved entanglement spectrum and the spatially-resolved feature spectrum. In this work, we highlight the equivalence between the feature and the entanglement spectra and extend the tripartite equivalence to sectors within the feature spectrum.}
\label{fig:03}
\end{figure}

In this work, we rigorously establish a \textbf{tripartite equivalence} between three fundamental spectra in non-interacting topological systems: the Feature Spectrum, the Entanglement Spectrum, and the Wilson Loop Spectrum. First, we prove that the feature and the entanglement spectra (with a consistent partitioning) have identical nonzero eigenvalues via Sylvester's determinant theorem. This indicates that the two spectra possess the same topological information, while the feature spectrum encodes the entanglement between different sectors, such as the spin-up and spin-down states when $\hat{O}$ is chosen as the spin operator. Next, we demonstrate that spectral flows in the spatially resolved feature spectrum and the spatially resolved entanglement spectrum can be adiabatically connected to the winding in the Wilson loop spectrum, completing the tripartite equivalence between them (FIG.~\ref{fig:03}). We propose the nested feature spectrum and show such a tripartite equivalence can be straightforwardly extended to the entanglement and the Wilson loop spectra defined on sectors in the feature spectrum, providing a more precise and subtle understanding of bulk-boundary correspondence in both the feature and the entanglement spectra. Our work thus gives a more physical interpretation of ground state entanglement, and highlights feature-energy complementarity as a fundamental manifestation of the bulk-boundary correspondence in topological phases.

\section{Results}
\subsection{Revisit: feature spectrum topology}
\par
\begin{figure}[h]
\centering
\includegraphics[width=0.8\linewidth]{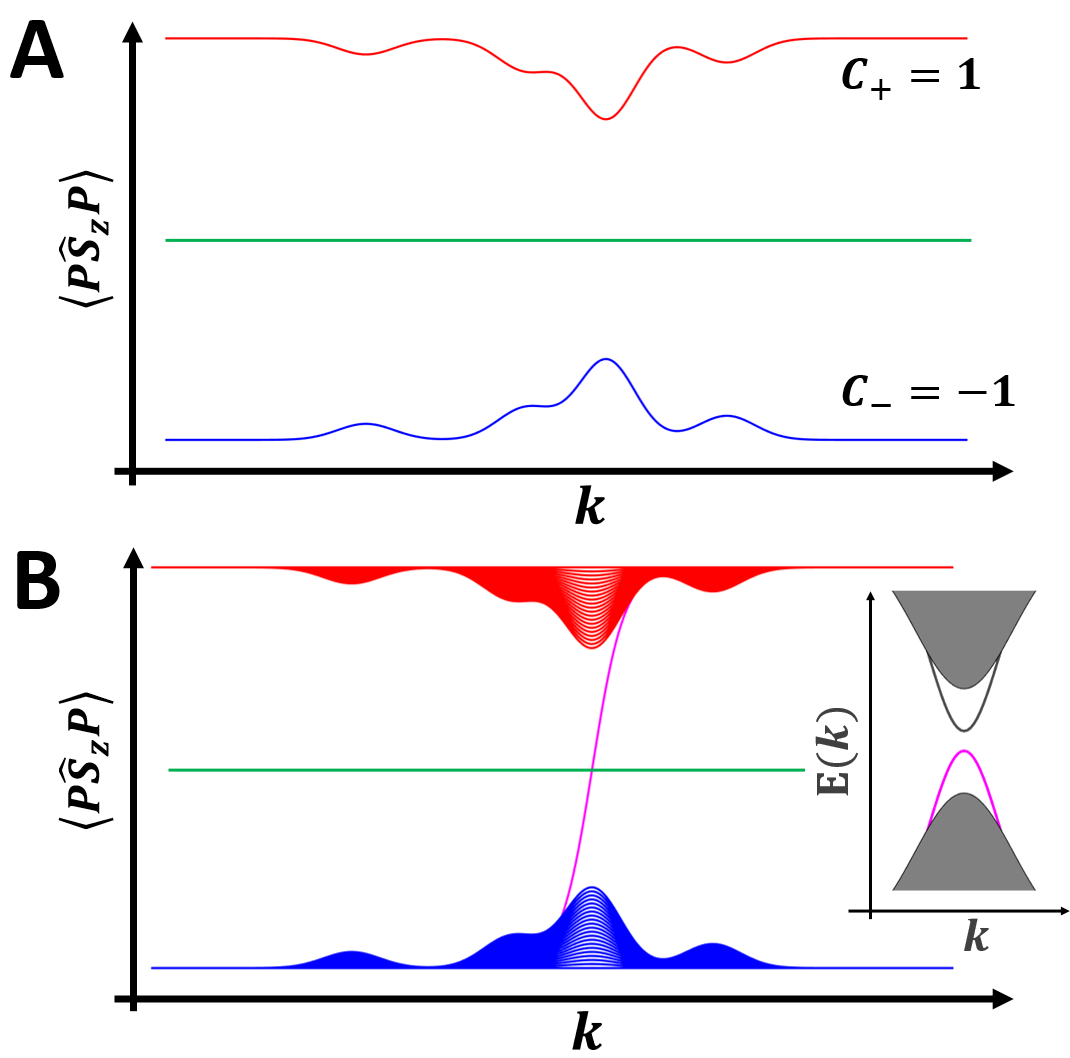}
\caption{(\textbf{A}) The schematic feature spectrum with feature $\hat{O} = \hat{S}_z$ of a spin-Chern insulator when spin-$U(1)$ symmetry is broken, leading to two sectors determined by the sign of their dispersion in the feature spectrum.
(\textbf{B}) The schematic of the corresponding feature energy complementarity, where the gapless spectral flows (magenta line) simultaneously appear in the feature spectrum when the energy edge states are gapped.}
\label{fig:01}
\end{figure}

 In the context of feature spectrum topology, the following projected operator is the main focus:
\begin{equation}\label{eq:01}
    \mathbf{F}=\hat{P}_{\text{(occ.)}}\hat{O}\hat{P}_{\text{(occ.)}},
\end{equation}
where $\hat{P}_{\text{(occ.)}}$ is the projection operator to the occupied states and $\hat{O}$ is the quantum number of interest, such as spin or orbital angular momentum. $\mathbf{F}$ is defined as the \emph{feature operator}, with a corresponding \emph{feature spectrum}. We assume $\hat{O}$ is translationally-invariant, which allows us to apply Bloch's theorem and define a band topology for the feature spectrum \cite{FeatureSpectrumTopology}. In addition, the occupied subspace can be further grouped into sectors based on well-separated bands in the feature spectrum. A schematic is shown in FIG.~\ref{fig:01}(A). When symmetry-breaking perturbations gap the boundary energy spectrum, the band topology of the feature spectrum allows us to define a bulk-boundary correspondence complementary to the energy spectrum. This \emph{feature-energy complementarity} is the key breakthrough of our framework, and a schematic is shown in FIG.~\ref{fig:01}(B).

\subsection{Revisit: the entanglement spectrum}\label{sec:02}
\par Next, we revisit the entanglement spectrum, which measures correlations between different partitions of the ground state, e.g., between regions A and B of a system. Non-trivial topology in a system shows up as gapless spectral flows in the entanglement spectrum, indicating a finite entanglement entropy. For non-interacting fermions, the entanglement Hamiltonian, $H^T$, is related to the single-particle correlation function, $C_A$, by $H^T = \ln(1 - C_A)/C_A$. Hence, in practice, only the eigenvalues of the single-particle correlation function are needed for the analysis of the entanglement spectrum \cite{PhysRevB.83.245132, brzezinska_2018, PhysRevLett.113.106801, J.Phys.A:Math.Gen.36.L205, PhysRevLett.105.115501, PhysRevB.84.195103, PhysRevB.87.035119}. For convenience, we also term the spectrum of the single-particle correlation function the entanglement spectrum. 

\par Various quantum operators that probe the electron's different internal properties have been suggested in the literature as suitable partitions for studying the entanglement spectrum in topological insulators, including orbital angular momentum \cite{PhysRevLett.101.010504}, spin or pseudospin \cite{doi:10.7566/JPSJ.85.043706, PhysRevB.96.165139}. Since these partition choices respect translational symmetry, we can define the bulk topological number using the eigenstates of the entanglement Hamiltonian in each partition \cite{doi:10.7566/JPSJ.85.043706, PhysRevB.96.165139, doi:10.7566/JPSJ.83.113705, doi:10.7566/JPSJ.84.043703}. This conceptual similarity to feature-spectrum topology, therefore, motivates us to study their connection.

\subsection{Equivalence between the feature and the entanglement spectra}\label{sec:equivalence}
\par Now we focus on relating the entanglement spectrum to the feature spectrum. The main result is that the feature and entanglement spectra have identical nonzero eigenvalues, and thus encode the same topological information. We proved this using Sylvester’s determinant theorem. First, given a chosen feature $\hat{O}$, we express it in spectral representation as:
\begin{equation}\label{eq:02}
\hat{O} =  \sum_{A}O_A \hat{P}_A,
\end{equation}
where $O_A$ is the $A$th eigenvalue. $\hat{P}_A=\sum_a^{N_A}\ket{\xi_a}\bra{\xi_a}$ is the projection operator to the Hilbert subspace $\mathcal{H}_{O_A}$ spanned by the $a$th eigenstate, $\ket{\xi_a}$, of the feature spectrum with eigenvalue, $O_A$, and $N_A$ is the number of such eigenstates. 

The entanglement spectrum of non-interacting fermions is the spectrum of the single-particle correlation function \cite{PhysRevB.101.115140, doi:10.7566/JPSJ.83.113705, doi:10.7566/JPSJ.84.043703}:
\begin{equation}\label{eq:03}
\mathbf{\hat{C}_A} =  \hat{P}_A \hat{P}_{(\text{occ.})} \hat{P}_A.
\end{equation}
When $\hat{O}$ is a symmetry, the entanglement spectrum is dispersionless with the eigenvalues equal $0$ or $1$. The bands with their eigenvalues equal $1$ belong to the Hilbert subspace $\mathcal{H}_{O_A,(\text{occ.})}=\mathcal{H}_{(\text{occ.})}\cap \mathcal{H}_{O_A}$. The bands with their eigenvalue equal $0$ belong to the Hilbert subspace $\mathcal{H}_{O_A,(\text{unocc.})}=\mathcal{H}_{(\text{unocc.})}\cap \mathcal{H}_{O_A}$, as well as states that do not belong to the Hilbert subspace $\mathcal{H}_{O_A}$, i.e., its complement $\mathcal{H}_{O_{A^{(c)}}}$. Here, $\mathcal{H}_{(\text{(un)occ.})}$ is the Hilbert subspace spanned by the (un)occupied states. When the $\hat{O}$ symmetry is broken perturbatively, the flat bands are adiabatically deformed into dispersive bands, with the corresponding eigenvalues lying in the interval $[0,1]$. We refer to these bands as the $1$-sector and $0$-sector depending on their adiabatic origin. Still, the band topology remains unchanged as long as the entanglement spectrum remains gapped \cite{doi:10.7566/JPSJ.85.043706, PhysRevB.96.165139, doi:10.7566/JPSJ.83.113705, doi:10.7566/JPSJ.84.043703}. 

\begin{figure}[h]
\centering
\includegraphics[width=0.8\linewidth]{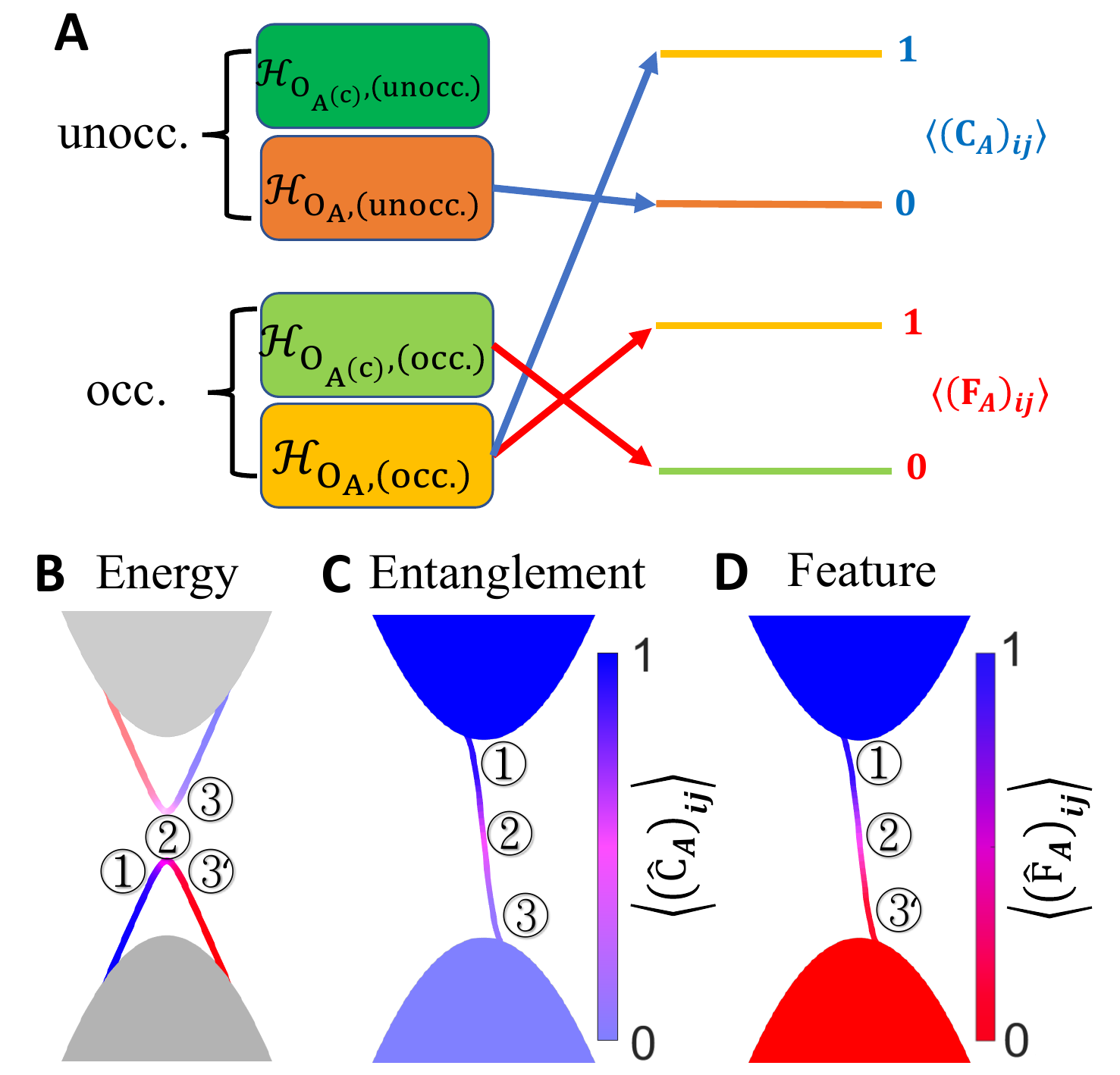}
\caption{(\textbf{A}) A schematic illustration of the entanglement spectrum from $(\mathbf{\hat{C}_A})_{ij}$ and the feature spectrum from $(\mathbf{\hat{F}_A})_{ij}$. In both spectra, the $1$-sector represents the occupied subspace of partition $A$ ($\mathcal{H}_{O_A,\text{(occ.)}}$). The $0$-sector in the former corresponds to unoccupied states in $A$ ($\mathcal{H}_{O_A,\text{(unocc.)}}$), while the one in the latter corresponds to occupied states in complementary partitions $A^{(c)}$ ($\mathcal{H}_{O_{A^{(c)}},\text{(occ.)}}$). 
(\textbf{B}) Under symmetry-breaking perturbations, even though boundary states become gapped in the energy spectrum, (\textbf{C}) the entanglement spectrum and (\textbf{D}) the feature spectrum still retain gapless spectral flows. The flows marked by \textcircled{1}–\textcircled{3}(\textcircled{3'}) track the evolution of boundary states. The color map shows the texture of the chosen feature, with darker (brighter) colors corresponding to occupied (unoccupied) states.}
\label{fig:02}
\end{figure}

The feature spectrum is the spectrum of the feature operator: 
\begin{equation}\label{eq:05}
\mathbf{\hat{F}_A} = \hat{P}_{(\text{occ.})}\hat{P}_A\hat{P}_{(\text{occ.})}.
\end{equation}
Similarly, the feature spectrum is dispersionless with eigenvalues equal to $0$ or $1$ when $\hat{O}$ is a symmetry. When $\hat{O}$ is not a symmetry, the bands become dispersive and are termed the $1$-sector and $0$-sector depending on their adiabatic origin. States in the $1$-sector belong to the Hilbert subspace $\mathcal{H}_{O_A,(\text{occ.})}=\mathcal{H}_{(\text{occ.})}\cap \mathcal{H}_{O_A}$. In contrast to the entanglement spectrum, states in the $0$-sector correspond to occupied states in the complement $\mathcal{H}_{O_{A^{(c)}}}$, i.e., $\mathcal{H}_{O_{A^{(c)}},(\text{occ.})}=\mathcal{H}_{(\text{occ.})}\cap \mathcal{H}_{O_{A^{(c)}}}$, as well as all of the unoccupied states $\mathcal{H}_{(\text{unocc.})}$. The similarities and differences between the feature and the entanglement spectra are illustrated in FIG.~\ref{fig:02}.

\par We now prove that the nonzero eigenvalues of the feature and entanglement spectra (c.f. Eqs.~\eqref{eq:03} and \eqref{eq:05}) are the same. First, explicitly writing out the matrix elements of $\mathbf{\hat{C}_A}$ and $\mathbf{\hat{F}_A}$  clarifies the different Hilbert subspaces in which they act:
\begin{widetext}
\begin{equation}\label{eq:06}
\begin{split}
    \mathbf{\hat{C}_A} & = \hat{P}_A\hat{P}_{(\text{occ.})}\hat{P}_A = \sum_{a,b=1}^{N_A}\ket{\xi_a}\bra{\xi_a}(\sum_{i=1}^{N_{(\text{occ.})}}\ket{\psi_i}\bra{\psi_i})\ket{\xi_b}\bra{\xi_b}
    \equiv \sum_{a,b=1}^{N_A}\ket{\xi_a}\bra{\xi_b}(\mathbf{\hat{C}_A})_{ab},
    \\ \mathbf{\hat{F}_A} & = \hat{P}_{(\text{occ.})}\hat{P}_A\hat{P}_{(\text{occ.})} = \sum_{i,j=1}^{N_{(\text{occ.})}}\ket{\psi_i}\bra{\psi_i}(\sum_{a=1}^{N_A}\ket{\xi_a}\bra{\xi_a})\ket{\psi_j}\bra{\psi_j}
    \equiv \sum_{i,j=1}^{N_{(\text{occ.})}}\ket{\psi_i}\bra{\psi_j}(\mathbf{\hat{F}_A})_{ij}.
\end{split}
\end{equation}
\end{widetext}
Here, $\ket{\psi_i}$ and $\ket{\xi_a}$ are eigenstates of $\hat{P}_{(\text{occ.})}$ and $\hat{O}$, respectively, and $N_{(\text{occ.})}$ denotes the number of occupied states. The matrix elements of $\mathbf{\hat{C}_A}$ and $\mathbf{\hat{F}_A}$ are denoted by $(\mathbf{\hat{C}_A})_{ij}$ and $(\mathbf{\hat{F}_A})_{ij}$, respectively. When $\hat{O}$ is a symmetry, the zero-eigenvalues of $(\mathbf{\hat{C}_A})_{ij}$ and $(\mathbf{\hat{F}_A})_{ij}$ will only include states in the Hilbert subspaces $\mathcal{H}_{O_A,(\text{unocc.})}$ and $\mathcal{H}_{O_{A^{(c)}},(\text{occ.})}$ respectively, as projection by $\hat{P}_A$ and $\hat{P}_{occ}$ discards  $\mathcal{H}_{A^{(\text{c})}}$ and $\mathcal{H}_{{\text{unocc.}}}$.  

The relation between $(\mathbf{\hat{C}_A})_{ij}$ and $(\mathbf{\hat{F}_A})_{ij}$ is clearly illustrated in (FIG.~\ref{fig:02}(A.)). The feature and entanglement spectra are closely related, as they can be expressed as:
\begin{equation}\label{eq:08}
\begin{split}
    (\mathbf{\hat{C}_A})_{ij} & = (\mathbf{UU^\dagger})_{ij},
    \\ (\mathbf{\hat{F}_A})_{ij} & = (\mathbf{U^\dagger U})_{ij},
\end{split}
\end{equation}
where $(\mathbf{U})_{ij}=\braket{\xi_i | \psi_j}$ is a $N_A\times N_{(\text{occ.})}$ matrix. In general, since $N_A\neq N_{(\text{occ.})}$ and $\mathbf{dim}((\mathbf{\hat{C}_A})_{ij})\neq \mathbf{dim}((\mathbf{\hat{F}_A})_{ij})$, the spectra of $(\mathbf{\hat{C}_A})_{ij}$ and $(\mathbf{\hat{F}_A})_{ij}$ are not identical. However, according to Sylvester's determinant theorem \cite{pozrikidis2014introduction}, the nonzero eigenvalues of both spectra in Eq.~\eqref{eq:08} are equal. That is, the spectra of $(\mathbf{\hat{C}_A})_{ij}$ and $(\mathbf{\hat{F}_A})_{ij}$ differ only in the $0$-sectors. This is consistent with the aforementioned physical meaning of states in the $1$- and $0$-sectors in both spectra when $\hat{O}$ is a symmetry. The states in the $1$-sectors in both the entanglement and the feature spectra belong to the same Hilbert subspace $\mathcal{H}_{O_A,(\text{occ.})}$. The difference between them lies in their respective $0$-sectors: the $0$-sector of the entanglement spectrum ($(\mathbf{\hat{C}_A})_{ij}$) is $\mathcal{H}_{O_A,(\text{unocc.})}$, whereas it is $\mathcal{H}_{O_{A^{(c)}},(\text{occ.})}$ for the feature spectrum ($(\mathbf{\hat{F}_A})_{ij}$) instead.  

\par We now show that the $1$-sectors of both spectra carry the same topological information as long as they are gapped. To define the band topology of these sectors, we consider only translationally invariant $\hat{O}$. When $\hat{O}$ is a symmetry, the $1$-sectors in both spectra differ only by a unitary transformation since they belong to the same Hilbert subspace $\mathcal{H}_{O_A,(\text{occ.})}$ (FIG.~\ref{fig:02}(A.)). The band topology of these $1$-sectors are thus the same because a unitary transformation does not affect topological invariants \cite{PhysRevB.89.155114, PhysRevB.84.075119, PhysRevB.95.075146} or the trace of the Berry curvature tensor \cite{EGUCHI1980213}. When $\hat{O}$-symmetry is broken perturbatively, the Hilbert subspaces spanned by the $1$-sector in both spectra are no longer identical, as they are now a mixture of states in the original $1$- and $0$-sectors in each spectrum. However, as long as these spectra remain gapped, their $1$-sectors can be adiabatically deformed back to the $\hat{O}$-symmetric cases, thus preserving the band topology of the $1$-sectors in the symmetry-breaking phases. 

\par We emphasize that the bulk-boundary correspondence for the band topology of these $1$-sectors has different meanings in each spectrum, as they connect different Hilbert subspaces. More specifically, the gapless spectral flow in the entanglement reflects how a chosen feature evolves from the occupied to unoccupied boundary states (FIG.~\ref{fig:02}(B) and (C)). The one in the feature spectrum reflects how a chosen feature evolves from a particular sector to its complementary sector, \emph{whilst remaining} within the occupied boundary states (FIG.~\ref{fig:02}(B) and (D)).

\subsection{Tripartite equivalence: the Wilson loop spectrum, the entanglement spectrum, and the feature spectrum}\label{sec:tripartite}
\par In this section, we prove a tripartite equivalence between the Wilson-loop spectrum,  the entanglement spectrum, and the feature spectrum. This is done by applying Sylvester’s determinant theorem to establish equivalence between the latter two in the real space, and then adiabatically connecting them to the Wilson-loop spectrum. First, we extend the feature and the entanglement spectra to real space. This leads to the spatially-resolved entanglement spectrum, which is a widely used method to predict topological boundary states. It is defined by a partial trace over a subsystem of the entire system in its spatial degree of freedom. For example, a spatial cut into the left and right subsystems and a partial trace over the right part of the subsystem. To be simple, we call the spatial region of interest, region $A$, and the region being partially traced, region $A_c$. For non-interacting fermions, the corresponding single-particle correlation function is formulated by \cite{PhysRevLett.96.110405, PhysRevB.101.115140, PhysRevB.83.245132, brzezinska_2018, PhysRevLett.113.106801, PhysRevLett.104.130502, PhysRevB.87.035119, Lin2024, PhysRevLett.107.036601, PhysRevLett.105.115501, PhysRevB.84.195103, PhysRevB.91.085119}:
\begin{equation}\label{eq:10}
    \mathbf{\hat{C}_A}(\vec{k}_\parallel,r_\perp) = \hat{P}_A\hat{P}_{(\text{occ.})}(\vec{k}_\parallel,r_\perp)\hat{P}_A,
\end{equation}
where the bands with eigenvalues of $1$ indicate the occupied states localized in the region $A$ and the bands with eigenvalues of $0$ indicate the unoccupied states localized in the region $A$ and states localized in the region $A_c$. Here, $r_\perp$ and $\vec{k}_{\parallel}$ are the components of $\vec{r}$ and $\vec{k}$ perpendicular and parallel to the cut direction, respectively. The projection operators $\hat{P}_A = \sum_{\vec{r}_i\in\,A}\ket{\phi_i}\bra{\phi_i}$ and $\hat{P}_{(\text{occ.})}(\vec{k}_\parallel,r_\perp) \equiv \sum_{k_\perp} e^{-ik_\perp\cdot r_\perp}\hat{P}_{(\text{occ.})}(\vec{k})e^{ik_\perp\cdot r_\perp}$, which has the matrix element $(\hat{P}_{(\text{occ.})}(\vec{k}_\parallel,r_\perp))_{ij} = \sum_{k_\perp} e^{-ik_\perp\cdot (r_i-r_j)}\braket{\phi_i|\hat{P}_{(\text{occ.})}(\vec{k})|\phi_j}$. Here, $\vec{r}_i,\vec{r}_j$ are the positions of orbitals $\ket{\phi_i},\ket{\phi_j}$ localized in the region $A$.

\par Correspondingly, we consider the spatially-resolved feature operator, 
\begin{equation}
\mathbf{\hat{F}_A}(\vec{k}_\parallel,r_\perp)=\hat{P}_{(\text{occ.})}(\vec{k}_\parallel,r_\perp)\hat{P}_A\hat{P}_{(\text{occ.})}(\vec{k}_\parallel,r_\perp), 
\end{equation}
which separates the occupied subspace into regions $A$ and $A_c$. The bands with eigenvalues equal to $1$ indicate the occupied states localized in the region $A$, and those equal to $0$ indicate the occupied states localized in the region $A_c$ and unoccupied states. Following the discussion in Sec.~\ref{sec:equivalence}, the spatially-resolved feature and the spatially-resolved entanglement spectra have identical nonzero eigenvalues. More specifically, in this case, Sylvester's determinant theorem is applied by expressing their matrix elements as 
\begin{equation}
\begin{split}
\left[\mathbf{\hat{F}_A}(\vec{k}_\parallel,r_\perp)\right]_{ij} & = \left[U(\vec{k}_\parallel,r_\perp)U^\dagger(\vec{k}_\parallel,r_\perp)\right]_{ij},
\\ \left[\mathbf{\hat{C}_A}(\vec{k}_\parallel,r_\perp)\right]_{ij} & = \left[U^\dagger(\vec{k}_\parallel,r_\perp) U(\vec{k}_\parallel,r_\perp)\right]_{ij}.
\end{split}
\end{equation}
Here, $\left[U(\vec{k}_\parallel,r_\perp)\right]_{ij}\equiv \braket{\xi_i|\psi_j(\vec{k}_\parallel,r_\perp)}$, where $\ket{\xi_i}$ and $\ket{\psi_j(\vec{k}_\parallel,r_\perp)}$ are eigenstates of $\hat{P}_A$ and $\hat{P}_{(\text{occ.})}(\vec{k}_\parallel,r_\perp)$, respectively. Moreover, the spectra of $\mathbf{\hat{C}_A}(\vec{k}_\parallel,r_\perp)$ and $\mathbf{\hat{F}_A}(\vec{k}_\parallel,r_\perp)$ can be adiabatically connected to the Wilson loop spectrum \cite{PhysRevLett.107.036601, PhysRevB.91.085119, J.Phys.A:Math.Gen.36.L205}; see Sec.~\ref{sec:methods} for details. Therefore, the feature spectrum of $\mathbf{\hat{F}_A}(\vec{k}_\parallel,r_\perp)$, the entanglement spectrum of $\mathbf{\hat{C}_A}(\vec{k}_\parallel,r_\perp)$, and the Wilson loop spectrum are mutually equivalent in characterizing non-trivial band topology (FIG.~\ref{fig:03}). 

\subsection{Nested feature spectrum and the tripartite equivalence}
\par Next, we present another key result of this work: the spectral flows of the entanglement spectrum defined on a feature spectrum sector guarantee feature-energy complementarity. Although each sector of the feature spectrum admits a well-defined band topology and bulk-boundary correspondence, its entanglement structure remains unexplored, even though non-trivial band topology is known to imply non-trivial entanglement. To clarify this, we define the \emph{nested feature spectrum}, and prove the tripartite equivalence between the nested feature spectrum, the entanglement spectrum, and the Wilson loop spectrum defined on a feature spectrum sector.

\par
\begin{figure}[h]
\centering
\includegraphics[width=0.8\linewidth]{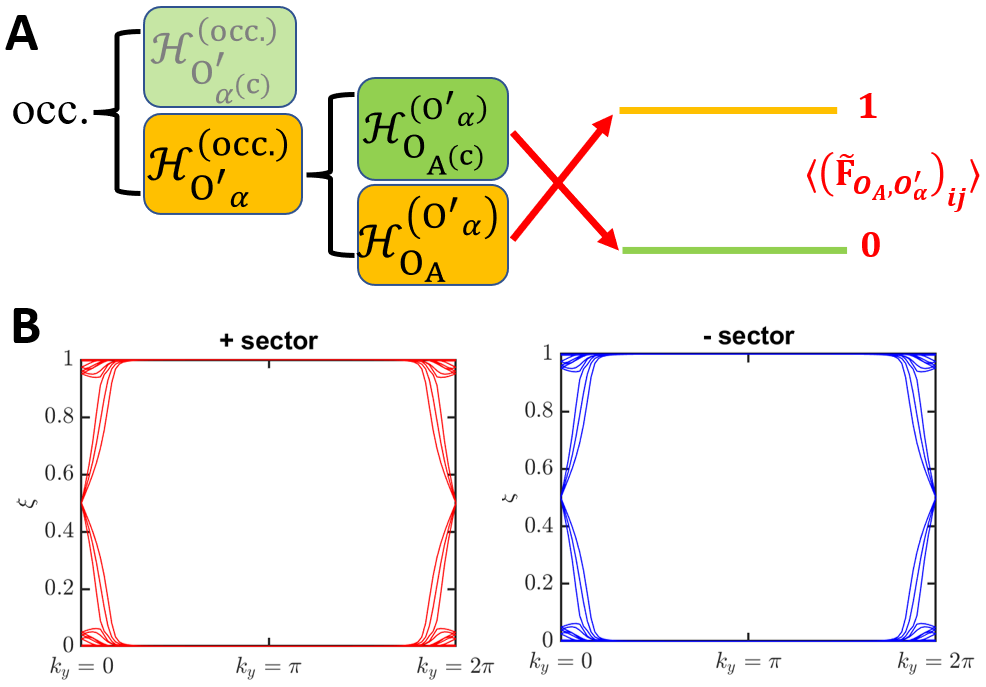}
\caption{(\textbf{A}) Schematic illustration of the nested feature spectrum derived from $(\mathbf{\tilde{F}_{O_A,O^{'}\alpha}})_{ij}$. The 1 (0) sector indicates the intersection of the occupied subspace associated with feature $\hat{O}'$ ($\mathcal{H}_{O^{'}\alpha}^{(\text{occ.})}$) and the 1 (0) sector of the feature spectrum defined by $\hat{O}$. These intersections define the subspaces $\mathcal{H}_{O_A}^{(O^{'}\alpha)}$ and $\mathcal{H}_{O_{A^{(c)}}}^{(O^{'}_\alpha)}$ for the 1- and 0-sectors, respectively.
(\textbf{B}) Feature-resolved spatial entanglement spectrum $\mathbf{\tilde{C}_{\hat{P}_A,O^{'}\alpha}}(\vec{k}\parallel,r_\perp)$ for the $+$ and $-$ sectors of the feature spectrum defined by $s_z\tau_x\sigma_0$ in a high-pseudospin-Chern insulator. Here, ${s_i}$, ${\tau_i}$, and ${\sigma_i}$ denote the spin, layer, and orbital degrees of freedom, respectively (Supplemental Materials Section~S1 for model details \cite{SM}).}
\label{fig:04}
\end{figure}

 Consider $\hat{O}$ and $\hat{O}'$ as two different internal quantum numbers of the electron. The nested feature spectrum of $\hat{O}$ defined within a sector of another feature spectrum of $\hat{O}^{'}$ is:
\begin{equation}\label{eq:nest_feature}
\begin{split}
\mathbf{\tilde{F}_{O_A,O^{'}_\alpha}} & = \hat{P}_{O',\alpha} \hat{P}_{A} \hat{P}_{O',\alpha}.
\end{split}
\end{equation}
Here, $\hat{P}_{O',\alpha}$ is the projection operator to the Hilbert subspace $\mathcal{H}_{O',\alpha}$ spanned by states in the $\alpha$-th sector of the feature spectrum of $\hat{O}'$. Similarly, $\hat{P}_{A}$ is the projection operator to the Hilbert subspace $\mathcal{H}_{O_A}$ spanned by the eigenstates of $\hat{O}$ with $A$-th eigenvalue $O_A$. The indices of $\mathbf{\tilde{F}_{O_A,O^{'}_\alpha}}$ denote the nested feature spectrum of $O_A$ defined within the $\alpha$-th sector of the $\hat{O}'$ feature spectrum. When $\hat{O}$ is a symmetry, states in the $1$- and $0$-sectors belong to the Hilbert subspace $\mathcal{H}_{O_A}^{(O^{'}\alpha)}=\mathcal{H}_{{O',\alpha}}\cap \mathcal{H}_{O_A}$ and $\mathcal{H}_{O_{A^{c}}}^{(O^{'}\alpha)}=\mathcal{H}_{{O',\alpha}}\cap \mathcal{H}_{O_{A^{(c)}}}$, respectively (FIG.~\ref{fig:04}(A)). $\mathcal{H}_{O_{A^{(c)}}}$ is the Hilbert subspace spanned by the eigenstates of $\hat{O}$ with eigenvalues other than $O_A$. 

\par The entanglement spectrum defined on a sector of a feature spectrum is the spectrum of a single-particle correlation function $\mathbf{\tilde{C}_{O_A,O^{'}_\alpha}}$:
\begin{equation}
\mathbf{\tilde{C}_{O_A,O^{'}_\alpha}}=\hat{P}_{A}\hat{P}_{O',\alpha}\hat{P}_{A}.
\end{equation}
For simplicity, we call $\mathbf{\tilde{C}_{O_A,O^{'}_\alpha}}$ the \emph{nested entanglement spectrum}. Since $\hat{P}_{O',\alpha}$ is a projection operator and $\hat{P}_{O',\alpha}^2=\hat{P}_{O',\alpha}$, we can express $\mathbf{\tilde{C}_{O_A,O^{'}_\alpha}}$ and $\mathbf{\tilde{F}_{O_A,O^{'}_\alpha}}$ as:
\begin{equation}\label{eq:14_new}
\begin{split}
    (\mathbf{\tilde{C}_{O_A,O^{'}_\alpha}})_{ij} & = (\mathbf{\tilde{U}\tilde{U}^\dagger})_{ij},
    \\ (\mathbf{\tilde{F}_{O_A,O^{'}_\alpha}})_{ij} & = (\mathbf{\tilde{U}^\dagger \tilde{U}})_{ij}.
\end{split}
\end{equation}
Here, $(\mathbf{\tilde{U}})_{ij}=\braket{\xi_i | \tilde{\psi}_j}$, and $\ket{\tilde{\psi}_j}$ and $\ket{\xi_i}$ are eigenstates of $ \hat{P}_{O',\alpha}$ and $\hat{P}_{A}$, respectively. Using an argument similar to the above discussion (c.f. Eq.~\eqref{eq:08}), the nested feature and the nested entanglement spectra are guaranteed to have the same nonzero eigenvalues by Sylvester's determinant theorem. Correspondingly, the $1$-sectors of both spectra belong to the same Hilbert subspace $\mathcal{H}_{O_A}^{(O^{'}\alpha)}$ when $\hat{O}$ is a symmetry. Furthermore, since $\hat{O}$ and $\hat{O}'$ are assumed to respect translational symmetry, this indicates that the $1$-sectors of $\mathbf{\tilde{C}_{O_A,O^{'}_\alpha}}$ and $\mathbf{\tilde{F}_{O_A,O^{'}_\alpha}}$ are topologically equivalent. The topological equivalence of the $1$-sectors in both spectra when $\hat{O}$ is not a symmetry can be derived in a similar manner using adiabatic continuity, as discussed in Sec.~\ref{sec:equivalence}.
\nocite{PhysRevLett.107.127205}

\par The tripartite equivalence can then be proven by applying these results to the spatially-resolved nested entanglement spectrum $\mathbf{\tilde{C}_{O_A,O^{'}_\alpha}}(\vec{k}_\parallel,r_\perp)$ and spatially-resolved nested feature spectrum $\mathbf{\tilde{F}_{O_A,O^{'}_\alpha}}(\vec{k}_\parallel,r_\perp)$, which are defined as:
\begin{equation}
\begin{split}
\mathbf{\tilde{C}_{O_A,O^{'}_\alpha}}(\vec{k}_\parallel,r_\perp) & = \hat{P}_A \hat{P}_{O',\alpha}(\vec{k}_\parallel,r_\perp) \hat{P}_A,
\\ \mathbf{\tilde{F}_{O_A,O^{'}_\alpha}}(\vec{k}_\parallel,r_\perp) & = \hat{P}_{O',\alpha}(\vec{k}_\parallel,r_\perp)\hat{P}_A \hat{P}_{O',\alpha}(\vec{k}_\parallel,r_\perp).
\end{split}
\end{equation}
Here, $\hat{P}_{O',\alpha}(\vec{k}_\parallel,r_\perp)\equiv \sum_{k_\perp} e^{-ik_\perp\cdot r_\perp} \hat{P}_{O',\alpha}(\vec{k})e^{ik_\perp\cdot r_\perp}$, and $\hat{P}_A$ is the projection operator to the spatial region of interest (region A). We explicitly write out the matrix element of $\mathbf{\tilde{F}_{O_A,O^{'}_\alpha}}(\vec{k}_\parallel,r_\perp)$ and $\mathbf{\tilde{C}_{O_A,O^{'}_\alpha}}(\vec{k}_\parallel,r_\perp)$ to demonstrate the application of Sylvester's determinant theorem in this case:
\begin{equation}
\begin{split}
\left[\mathbf{\tilde{F}_{O_A,O^{'}_\alpha}}(\vec{k}_\parallel,r_\perp)\right]_{ij} & = \left[\mathbf{\tilde{U}}(\vec{k}_\parallel,r_\perp)\mathbf{\tilde{U}}^\dagger(\vec{k}_\parallel,r_\perp)\right]_{ij},
\\ \left[\mathbf{\tilde{C}_{O_A,O^{'}_\alpha}}(\vec{k}_\parallel,r_\perp)\right]_{ij} & = \left[\mathbf{\tilde{U}}^\dagger(\vec{k}_\parallel,r_\perp) \mathbf{\tilde{U}}(\vec{k}_\parallel,r_\perp)\right]_{ij}.
\end{split}
\end{equation}
Here, $\mathbf{\tilde{U}}^\dagger(\vec{k}_\parallel,r_\perp)=\braket{\xi_i | \tilde{\psi}_j(\vec{k}_\parallel,r_\perp)}$, and $\ket{\tilde{\psi}_j(\vec{k}_\parallel,r_\perp)}$ and $\ket{\xi_i}$ are eigenstates of $ \hat{P}_{O',\alpha}(\vec{k}_\parallel,r_\perp)$ and $\hat{P}_{A}$, respectively.

\par In a similar manner to Sec.~\ref{sec:tripartite}, we can adiabatically connect the spatially-resolved nested entanglement spectrum $\mathbf{\tilde{C}_{O_A,O^{'}_\alpha}}(\vec{k}_\parallel,r_\perp)$ with the Wilson loop spectrum defined on that sector, with details of the proof included in Sec. ~\ref{sec:methods}. Since the Wilson loop defined on a feature spectrum sector directly reveals the existence of a gapless feature edge state, and we have established the equivalence between $\mathbf{\tilde{C}_{O_A,O^{'}_\alpha}}(\vec{k}_\parallel,r_\perp)$ and $\mathbf{\tilde{F}_{O_A,O^{'}_\alpha}}(\vec{k}_\parallel,r_\perp)$, the spectral flows of $\mathbf{\tilde{C}_{O_A,O^{'}_\alpha}}(\vec{k}_\parallel,r_\perp)$ thus guarantees feature-energy complementarity rather than gapless energy eigenstates at the boundary. This discussion thus completes the proof of the tripartite equivalence between the nested entanglement spectrum, the nested feature spectrum, and the Wilson loop spectrum defined on sectors of a feature spectrum.

\par Our theory is applicable to any translationally-invariant quantum number of the electron, and in demonstration, we present the spin-pseudospin resolved entanglement spectrum of the composite projection operator $\hat{P}_{O^{'}_\alpha=\pm}(\vec{k}_\parallel,r_\perp)$, for the spin-pseudospin feature operator $\hat{O}^{'}= s_z\tau_x\sigma_0$. The feature-resolved nested entanglement spectrum displays gapless spectral flows in topologically non-trivial sectors in the spin-pseudospin feature spectrum, as shown in FIG.~\ref{fig:04}(B). The special case of spin-resolved nested entanglement spectrum, $\mathbf{C_{\hat{P}_A,\hat{S}_z = \pm \frac{1}{2}}}(\vec{k}_\parallel,r_\perp) = \hat{P}_{\hat{S}_z=\pm\frac{1}{2}}\hat{P}_{(\text{occ.})}(\vec{k}_\parallel,r_\perp)\hat{P}_{\hat{S}_z=\pm\frac{1}{2}}$, and its relation to spin-resolved edge states has also been considered by another group \cite{Lin2024}.

\subsection{Experimental signature}
\par The boundary entanglement entropy can be probed indirectly in specific cases. Since the main contribution of the entanglement entropy is mainly due to the boundary states originating from the feature-energy complementarity \cite{brzezinska_2018, PhysRevB.73.245115, PhysRevB.83.245132, PhysRevB.87.035119}, probing the boundary entanglement spectrum, or equivalently, the boundary feature spectrum, is sufficient to understand the non-trivial ground state entanglement in the system. Our previous work \cite{FeatureSpectrumTopology} focusing on certain feature Chern insulators has demonstrated that the boundary feature spectrum can be probed by measuring the optical transition rate under exposure to monochromatic unpolarized light. In this setup, the feature Chern insulator is characterized by a feature operator $\hat{O}$ satisfying $\hat{O}^2=\pm1$, which matches the pseudospin of the low-energy states. Thus, utilizing the equivalence between the feature and the entanglement spectra, the boundary entanglement spectrum and its main contribution of the entanglement entropy can be inferred from this measurement.

\section{Discussion}
\par We have presented the feature spectrum framework, which enables us to investigate finer details of band topology. It does so by partitioning the ground-state Hilbert space into different sectors via a projection operator. Using Sylvester's theorem, we have proven that the feature and the entanglement spectra have identical non-zero eigenvalues, implying that states with non-zero eigenvalues carry the same band topology. Moreover, via adiabatic connection, we prove a tripartite equivalence between the feature, the entanglement, and the Wilson loop spectra. Such a tripartite equivalence is further extended to the entanglement and the Wilson loop spectra defined on a sector of the feature spectrum. This clearly shows that when the entanglement spectrum is restricted to a specific feature-spectrum sector and exhibits spectral flow, feature–energy complementarity is ensured; in other words, the boundary must display either gapless energy modes or feature-resolved eigenstates. Furthermore, this complementarity allows us to probe the entanglement entropy by optical measurements of the feature-resolved edge states. 

We note that a nontrivial topology of the entanglement spectrum does not necessarily guarantee the presence of gapless boundary states in the energy spectrum. Since we have proven that the topologies of the entanglement spectrum and the feature spectrum are equivalent, it is possible for topologically nontrivial boundary states to be gapped in the energy spectrum while remaining gapless in the entanglement or feature spectrum. Using the entanglement spectrum to classify band topology in interacting systems lies beyond the scope of the present work and may require further scrutiny, as the equivalence between both spectra relies on the equivalence of the single-particle correlation function and the entanglement spectrum. However, the latter equivalence holds only for non-interacting fermionic systems. Moreover, electron–electron interactions can generate new states or open additional energy gaps, which further complicates the analysis. Whether such entanglement-spectrum topology implies gapless energy boundary states, even when the symmetry is restored, remains an open question.

\par In the larger scheme of things, the feature spectrum provides a complementary perspective to the entanglement spectrum (Table~\ref{tab:01}), offering a unified and physically transparent probe of the non-trivial band topology associated with different local quantum numbers. By establishing the tripartite equivalence, our work clarifies the structure of ground-state entanglement in non-interacting fermionic systems beyond the conventional energy-band description. More broadly, this framework offers a systematic route to explore and characterize more realistic, symmetry-broken topological materials, in which topological and entanglement information is encoded in the feature spectrum eigenstates rather than in the energy eigenstates. These ideas can be readily extended to Floquet systems, in which periodic driving can alter symmetries and render the bulk–boundary correspondence based solely on energy eigenstates incomplete.

\begin{table*}[h]
\begin{tabular}{ || m{5em} | m{7.4cm} | m{7.4cm} || }
  \hline
   \quad & \textbf{Entanglement Spectrum} & \textbf{Feature Spectrum} \\ 
  \hline
  \textbf{Operator} & $\mathbf{\hat{C}_A}=\hat{P}_A\hat{P}_{(\text{occ.})}\hat{P}_A$ & $\mathbf{F} = \sum_AO_A\hat{P}_{(\text{occ.})}\hat{P}_A\hat{P}_{(\text{occ.})} = \sum_AO_A\mathbf{\hat{F}_A}$  \\ 
  \hline
  \textbf{Matrix Element} & \begin{equation*}(\mathbf{\hat{C}_A})_{ab}=\sum_{i=1}^{N_{(\text{occ.})}}\braket{\xi_a|\psi_i}\braket{\psi_i|\xi_b}\end{equation*} & \begin{equation*}(\mathbf{\hat{F}_A})_{ij}=\sum_{a=1}^{N_A}\braket{\psi_i|\xi_a}\braket{\xi_a|\psi_j}, i,j\leq N_{(\text{occ.})}\end{equation*} 
  \\
  \hline
  \textbf{$1$-sector} &
\multicolumn{2}{m{14.8cm}||}{\centering
The occupied part of the partition $A$, i.e., $\mathcal{H}_{O_A,(\text{occ.})}$.
}  \\
  \hline
  \textbf{$0$-sector} & The unoccupied part of the partition $A$, i.e., $\mathcal{H}_{O_A,(\text{unocc.})}$. & The occupied part of other sectors, i.e., $\mathcal{H}_{O_{A^{(c)}},(\text{occ.})}$. \\
  \hline
  \textbf{Property Detected} & Whether the ground state can be disentangled \cite{doi:10.7566/JPSJ.84.043703}. & Whether the occupation flows between sectors \cite{FeatureSpectrumTopology}. \\
  \hline
\end{tabular}
\caption{A summarized table for comparisons between feature-cut ES and feature spectrum with the feature $\hat{O}=\sum_AO_A\hat{P}_A$. $1(0)$-bands indicate the bands with eigenvalue $1(0)$ in the spectrum, where $\{\ket{\xi_a}\}$ and $\{\ket{\psi_i}\}$ are the eigenstates of $\hat{O}$ and energy eigenstates, respectively.}
\label{tab:01}
\end{table*}

\section{Methods}\label{sec:methods}
\subsection{Detailed proof of the tripartite equivalence}\label{sec:A}
\par In this section, we prove the equivalence between the entanglement and the Wilson loop spectra, as well as the equivalence between the feature and the Wilson loop spectra. First, continuing on Eq.~\eqref{eq:10}, we note that if we choose the region $A$ to be the region on the left-hand side of the cut and the region $A_c$ to the region on the right-hand side of the cut, the spectrum of $\mathbf{\hat{C}_A}(\vec{k}_\parallel,r_\perp)$ in Eq.~\eqref{eq:10} is $\hat{P}_{LL}$, and can be adiabatically deformed to $1-2\hat{P}_{LL}$, where the projection operator is written as $\hat{P}_{(\text{occ.})}=\begin{pmatrix}\hat{P}_{LL} & \hat{P}_{LR} \\ \hat{P}_{RL} & \hat{P}_{RR} \end{pmatrix}$ with $L(R)$ indicating the left-(right-)hand side of the cut \cite{PhysRevLett.107.036601}. The spectrum of $1-2\hat{P}_{LL}$ ensembles the occupied states outside the left-hand side of the cut with spectral value $1$ and the states inside with spectral value $-1$. On the other hand, the spectrum of $-\hat{P}_{LL}$ ensembles the occupied states outside the right-hand side of the cut with spectral value $0$ and the states inside with spectral value $-1$. Therefore, we can adiabatically deform the spectrum of $\hat{P}_{LL}$ to $1-2\hat{P}_{LL}$ up to a minus sign. Since the spectrum of $1-2\hat{P}_{RR/(LL)}$ is shown to be able to adiabatically connect to the spectrum of $\hat{P}_{(\text{occ.})}\hat{r}_\perp \hat{P}_{(\text{occ.})}$ through $\hat{P}_{(\text{occ.})}V_0\hat{P}_{(\text{occ.})}$ with $V_0=-\text{sign}(r_\perp)$, the spectrum of $\mathbf{\hat{C}_A}(\vec{k}_\parallel,r_\perp)$ can thus be adiabatically connected to the Wilson loop spectrum because the spectrum of $\hat{P}_{(\text{occ.})}\hat{r}_\perp \hat{P}_{(\text{occ.})}$ is equivalent to the Wilson loop spectrum \cite{PhysRevLett.107.036601}.

\par Next, we discuss the connection between the feature and the Wilson loop spectra. The feature spectrum of the feature operator $\mathbf{\hat{F}_A}(\vec{k}_\parallel,r_\perp)$ can be related to the Wilson loop spectrum in the same way. More specifically, $\hat{P}_{(\text{occ.})}V_0\hat{P}_{(\text{occ.})}=\hat{P}_{(\text{occ.})}(\hat{P}_A-\hat{P}_B)\hat{P}_{(\text{occ.})}=\mathbf{\hat{F}_A}(\vec{k}_\parallel,r_\perp)-\mathbf{F_B}(\vec{k}_\parallel,r_\perp)$ consists of two parts. The $1$-sector of the feature spectrum of $\mathbf{F_{A(B)}}(\vec{k}_\parallel,r_\perp)$ represents the occupied states localized on the left(right)-hand side of the cut while the $0$-sector represents the unoccupied states and the occupied states localized on another side of the cut. Since $\hat{P}_{(\text{occ.})}V_0\hat{P}_{(\text{occ.})}=\mathbf{\hat{F}_A}(\vec{k}_\parallel,r_\perp)-\mathbf{F_B}(\vec{k}_\parallel,r_\perp)$, its spectral flows originating from the $1$-sectors represent the occupation flowing from the left-hand side of the cut to the unoccupied states or to the occupied states localized on the right-hand side of the cut, which carries the same information as the spectral flows in the spectrum of $\mathbf{\hat{F}_A}(\vec{k}_\parallel,r_\perp)$. Therefore, the feature spectrum of the feature operator $\mathbf{\hat{F}_A}(\vec{k}_\parallel,r_\perp)$ can be related to the spectrum of $\hat{P}_{(\text{occ.})}V_0\hat{P}_{(\text{occ.})}$. Then, by using the relationship between the spectrum of $\hat{P}_{(\text{occ.})}V_0\hat{P}_{(\text{occ.})}$ and the spectrum of $\hat{P}_{(\text{occ.})}\hat{r}_\perp \hat{P}_{(\text{occ.})}$ and the equivalence between the spectrum of $\hat{P}_{(\text{occ.})}\hat{r}_\perp \hat{P}_{(\text{occ.})}$ with the Wilson loop spectrum \cite{PhysRevLett.107.036601}, the feature spectrum of the feature operator $\mathbf{\hat{F}_A}(\vec{k}_\parallel,r_\perp)$ can be related to the Wilson loop spectrum.

\par The above results can be easily extended to the nested entanglement spectrum and the nested feature spectrum by replacing $\hat{P}_{(\text{occ.})}$ with $\hat{P}_{O',\alpha}$ in the above argument. Here, $\hat{P}_{O',\alpha}$ is the projection operator to the $\alpha$-sector in another feature spectrum $\mathbf{F}=\hat{P}_{(\text{occ.})}\hat{O}'\hat{P}_{(\text{occ.})}$ with a translationally invariant feature $\hat{O}'$. First, the spatially resolved nested entanglement spectrum and feature spectrum are $\mathbf{\tilde{C}_{O_A,O^{'}_\alpha}}(\vec{k}_\parallel,r_\perp) = \hat{P}_A\hat{P}_{O',\alpha}(\vec{k}_\parallel,r_\perp)\hat{P}_A$ and $\mathbf{\tilde{F}_{O_A,O^{'}_\alpha}}(\vec{k}_\parallel,r_\perp) = \hat{P}_{O',\alpha}(\vec{k}_\parallel,r_\perp)\hat{P}_A\hat{P}_{O',\alpha}(\vec{k}_\parallel,r_\perp)$. The connection between $\mathbf{\tilde{C}_{O_A,O^{'}_\alpha}}(\vec{k}_\parallel,r_\perp)$ and the Wilson loop spectrum of a feature sector can be demonstrated as follows. Since $\mathbf{\tilde{C}_{O_A,O^{'}_\alpha}}(\vec{k}_\parallel,r_\perp)$ is $\hat{P}_{\alpha,LL}$, where $\hat{P}_{O',\alpha}=\begin{pmatrix}\hat{P}_{\alpha,LL} & \hat{P}_{\alpha,LR} \\ \hat{P}_{\alpha,RL} & \hat{P}_{\alpha,RR} \end{pmatrix}$ with $L(R)$ indicating the left (right)-hand side of the cut, its spectrum can be adiabatically connected to the spectrum of $\hat{P}_{O',\alpha}\hat{r}_\perp \hat{P}_{O',\alpha}$ through $\hat{P}_{O',\alpha}V_0\hat{P}_{O',\alpha}$. This adiabatic connection holds because the spectrum of $\hat{P}_{\alpha,LL}$ can be adiabatically connected to the spectrum of $\hat{P}_{(\text{occ.})}-2\hat{P}_{\alpha,LL}$ up to a minus sign in the occupied subspace. More specifically, since $\hat{P}_{O',\alpha} V_0 \hat{P}_{O',\alpha} + (\hat{P}_{(\text{occ.})}-\hat{P}_{O',\alpha})V_0(\hat{P}_{(\text{occ.})}-\hat{P}_{O',\alpha}) = \begin{pmatrix}\hat{P}_{(\text{occ.})}-2\hat{P}_{\alpha,LL} & 0 \\ 0 & \hat{P}_{(\text{occ.})}-2\hat{P}_{\alpha,RR} \end{pmatrix}$, with the assumption of no states in the $\alpha$-sector localized exactly on one side of the cut, the spectra of $\hat{P}_{O',\alpha} V_0 \hat{P}_{O',\alpha}$ and $\hat{P}_{(\text{occ.})}-2\hat{P}_{\alpha,LL}$ are identical \cite{PhysRevLett.107.036601}. Therefore, $\mathbf{\tilde{C}_{O_A,O^{'}_\alpha}}(\vec{k}_\parallel,r_\perp)$ is equivalent to the Wilson loop spectrum of a feature sector since the latter is equivalent to the spectrum of $\hat{P}_{O',\alpha}\hat{r}_\perp \hat{P}_{O',\alpha}$, which further connects to the spectrum of $\hat{P}_{O',\alpha} V_0 \hat{P}_{O',\alpha}$ \cite{FeatureSpectrumTopology}. We can establish the equivalence of the feature spectrum $\mathbf{\tilde{F}_{O_A,O^{'}_\alpha}}(\vec{k}_\parallel,r_\perp)$ of a feature sector to the Wilson loop spectrum of a feature sector using a similar technique by identifying $\hat{P}_{O',\alpha} V_0 \hat{P}_{O',\alpha} = \mathbf{\tilde{F}_{O_A,O^{'}_\alpha}}(\vec{k}_\parallel,r_\perp)-\mathbf{\tilde{F}_{\hat{P}_B,O^{'}_\alpha}}(\vec{k}_\parallel,r_\perp)$ and following the arguments in the previous paragraph.


\section*{Data availability}
\par The data that support the findings of this article are not publicly available upon publication because it is not technically feasible and/or the cost of preparing, depositing, and hosting the data would be prohibitive within the terms of this research project. The data are available from the authors upon reasonable request.

\section*{Code availability}
\par The source code of the quantum vacuum solver is available from the corresponding author on reasonable request.


\section*{Author Contributions}
Y.-C. H. developed the initial concept. Y.-C. H. and T. O. performed the mathematical derivations. Y.-C. H. constructed the tight-binding models and computed the corresponding results. H. L. supervised the project. Y.-C. H., T. O., and H. L. discussed the paper together. Y.-C. H. and T. O. wrote the paper together.

\bibliography{apssamp}

\end{document}